\documentclass[12pt,a4paper]{article}
\usepackage{amsmath}
\usepackage{amsfonts}
\usepackage{amssymb,multirow,multicol,booktabs}
\usepackage{graphicx,hyperref,float}
\hypersetup{colorlinks = true, linkcolor = blue, anchorcolor = blue, citecolor = blue, filecolor = blue, urlcolor = blue}
\usepackage[top=1.05in,left=1in,right=1in,bottom=1.05in]{geometry}
\newcommand\BibTeX{{\rmfamily B\kern-.05em \textsc{i\kern-.025em b}\kern-.08em
T\kern-.1667em\lower.7ex\hbox{E}\kern-.125emX}}

\begin{document}
	
%\runtitle{Stability Selection for Lasso, Ridge and Elastic Net}

\small{
\begin{center}
		\textbf{\Large Stability Selection for Lasso, Ridge and Elastic Net Implemented with AFT Models}\\	\vspace{.5cm}

        \textbf{Running Title: Stability Selection for AFT Models}
\end{center}
	\vspace{.7cm}
	
	\begin{center}
		{Md Hasinur Rahaman Khan\footnote{the corresponding author}, Institute of Statistical Research and Training, University of Dhaka, Dhaka-1000, Bangladesh, Email: hasinur@isrt.ac.bd}\\
%Phone:+8801725106661, Fax:(880-2) 8615583
		{Anamika Bhadra, Institute of Statistical Research and Training, University of Dhaka, Dhaka-1000, Bangladesh, Email: abhadra@isrt.ac.bd}\\and\\
		{Tamanna Howlader, Institute of Statistical Research and Training, University of Dhaka, Dhaka-1000, Bangladesh, Email: tamanna@isrt.ac.bd}

%		E-mail: $^1$bhossain@isrt.ac.bd, $^2$hasinur@isrt.ac.bd}\\
		\vspace{0.5cm}
		
		\noindent\rule{16cm}{0.4pt}
	\end{center}
%
%\textbf{\large{Total Word Counts: 4100}}\\
%

\textbf{\large{Abstract}}\\

%The instability in the selection of models is a major concern with data sets containing a large number of covariates. We focus on stability selection which is used as a technique to improve variable selection performance for a range of selection methods, based on aggregating the results of applying a selection procedure to sub-samples of the data where the observations are subject to right censoring. The accelerated failure time (AFT) models have proved useful in many contexts, though heavy censoring (as for example in cancer survival) and high dimensionality (as for example in micro-array data) cause difficulties for model fitting and model selection. We implement the stability selection approach using three variable selection techniques---Lasso, ridge regression, and elastic net applied to censored data using AFT models. We compare the performances of these regularized techniques with and without stability selection approaches with simulation studies and a breast cancer data analysis. The results suggest that stability selection gives always stable scenario about the selection of variables and that as the dimension of data increases the performance of methods with stability selection also increases than the methods without stability selection irrespective of the collinearity between the covariates.\\
The instability in the selection of models is a major concern with
data sets containing a large number of covariates. We focus on
stability selection which is used as a technique to improve
variable selection performance for a range of selection methods,
based on aggregating the results of applying a selection procedure
to sub-samples of the data where the observations are subject to
right censoring. The accelerated failure time (AFT) models have
proved useful in many contexts including the heavy censoring (as for
example in cancer survival) and the high dimensionality (as for
example in micro-array data). We implement the stability selection approach
using three variable selection techniques---Lasso, ridge
regression, and elastic net applied to censored data using AFT
models. We compare the performances of these regularized
techniques with and without stability selection approaches with
simulation studies and a breast cancer data analysis. The results
suggest that stability selection gives always stable scenario
about the selection of variables and that as the dimension of data
increases the performance of methods with stability selection also
improves compared to methods without stability selection
irrespective of the collinearity between the covariates.\\

	}\\
\noindent\rule{16cm}{0.4pt}
%\newpage

\section{Introduction}
The problem of variable selection for best predictive accuracy has
received a huge amount of attention over the last 15 years,
motivated by the desire to understand structure in massive data
sets that are now routinely encountered across many scientific
disciplines. In molecular biology for instance, microarray
experiments are being used to record expression measurements for
thousands of genes simultaneously and it is of interest to
identify a small subset of genes that influence disease prognosis
or survival. This paper is concerned with variable selection for
high dimensional survival data in which the number of covariates
($p$) is large compared to the number of replications or sample
size ($n$). Standard survival regression techniques are not
amenable to such data and thus current research centers on
adapting these methods to the large $p$ and small $n$ scenario.

Much of the earlier work on variable selection for linear
regression models has been extended to survival regression models.
Examples include best subset selection, stepwise selection,
bootstrap procedures \cite{ref:Sauerbrei}, Bayesian variable
selection [\cite{ref:Faraggi}, \cite{ref:Ibrahim}] and popular
penalization methods such as the Lasso (least absolute shrinkage
and selection operator) \cite{tibshirani1996regression}, ridge
regression \cite{hoerl1970ridge}, least angle regression selection
(LARS) \cite{efron2004least}, the elastic net
\cite{zou2005regularization}, the Dantzig selector
\cite{candes2007dantzig} and Double Dantzig
\cite{james2009generalized}. Penalization methods put penalties on
the regression coefficients. By properly balancing goodness of fit
and model complexity, penalization approaches can lead to
parsimonious models with reasonable fit. However, instability in
the selection approaches has been encountered in the context of
linear regression. For example, it has been shown that in general
the lasso is not variable selection consistent \cite{ref:Leng}.

The Cox proportional hazards regression model is one of the most
widely used regression models for analyzing censored survival
data. Several methods have been proposed for variable selection
under this model. For instance, the Lasso has been used in gene
expression analysis with survival data in
\cite{tibshirani1997Lasso}, \cite{ref:Gui_a} and
\cite{ref:Ternès}, the smoothly clipped absolute deviation penalty
(SCAD) has been used in \cite{ref:Fan}, the adaptive Lasso penalty
in \cite{ref:Zhang}, the kernel transformation in \cite{ref:Li},
and threshold gradient descent minimization in \cite{ref:Gui_b}.
An alternative to the Cox regression model for the analysis of
censored survival data is the accelerated failure time model
(AFT). The AFT model is a linear regression model in which
logarithm of the failure time is directly regressed on covariates
so that it is intuitively more interpretable than the Cox model
\cite{wei1992accelerated}. The AFT model generates a summary
measure that is interpreted in terms of the survival curve instead
of the hazard ratio and this makes it particulary useful for
certain applications such as experimental aging research
\cite{ref:Swindell}. Unlike the Cox model, there have been only a
limited number of studies focussing on variable selection for the
AFT model. See for example, \cite{ref:Huang1}, \cite{ref:Huang2},
\cite{ref:Wang}. However, heavy censoring and high dimensionality
are known to cause instabilities in variable or model selection.

Recently, Meinshausen and B{\"u}hlmann
\cite{meinshausen2010stability} have proposed stability selection
which is based on subsampling to obtain more stable selection of
the variables. The basic idea is that instead of applying a
regularization algorithm to the whole data set to determine the
selected set of variables, one instead applies it several times to
random subsamples of the data of size $[\frac{n}{2}]$ and chooses
those variables that are selected most frequently on the
subsamples. Stability selection applied to the linear regression
model is attractive for a number of reasons. First, the method is
extremely general and therefore has wide applicability. Second,
with stability selection, results are much less sensitive to the
choice of the regularization. This is a real advantage in
high-dimensional problems with $ p >> n $, as it is very hard to
estimate the noise level in these settings. Thirdly, stability
selection makes the regularization technique variable selection
consistent in settings where the original methods fail. The
purpose of this study is to examine the effect of stability
selection \cite{meinshausen2010stability} on variable selection in
case of the AFT model using three widely used techniques in the
literature, namely, the Lasso \cite{tibshirani1996regression},
ridge regression \cite{hoerl1970ridge} and elastic net
\cite{zou2005regularization} methods. Comparisons are made with
and without stability selection for both low-dimensional and
high-dimensional right censored data using simulation studies and
a breast cancer data set. The results will provide insights on
whether stability selection is a viable strategy for improving
variable selection when analyzing high dimensional survival data.

The paper is organized as follows. Section 2 presents the AFT
model and the weighted least square estimation method
\cite{stute1993consistent}, which accounts for censoring.
Application of the Lasso, ridge and elastic net penalties to the
AFT model estimation are described. Steps of the stability
selection \cite{meinshausen2010stability} procedure are also
summarized in this section. In Section 3, details are presented
for a simulation study to evaluate the performance of stability
selection in improving stability of covariate selection in low
dimensional and high dimensional settings. In Section 4, an
application to a gene expression data in breast cancer is
described. The findings of the study are discussed in Section 5.

%However the aim is not to address which variable selection is superior, rather we compare the performance of stability selection for these three methods using statistical criterion: the selection error rate, which allows to quantify the prediction performance of the final model selected.
%
%The rest of the study organised as follows. In section 2 we define the regularized framework of SWLS and variable selection methods for censored data setting.
%In section 3 we use simulated data to compare the performance of  three modified variable selection approaches: Lasso, ridge regression and elastic net for
%censored data due to stability selection. In section 4 real data  analysis is provided. Discussion and concluding remarks are given in the last Section.
\subsection{Accelerated failure time (AFT) model}
The AFT model is a linear regression model in which the response
variable is the logarithm of the failure time $ T_{i}$
\cite{kalbfleisch2011statistical}. Let $ Y_{i} $ be the logarithm
of failure time and $ X_{i} $ be the covariate vector for $ i^{th}
$ individual in a random sample of size $ n $. Also let $ Y_{(1)}
\le \cdots \le Y_{(n)} $ be the ordered logarithm of survival
times, and $ \delta_{(1)}, \cdots , \delta_{(n)} $ the associated
censoring indicators. Then the AFT model is defined by

\begin{equation} \label{eq:aft}
Y_{i} = \beta_{0} + X_{i}^{T}\beta + \varepsilon_{i},
~\,\,\,\,\,\, i = 1, \cdots, n,
\end{equation}
where $ \beta_{0} $ is the intercept term, $ \beta $ is the
unknown $ p \times 1 $ vector of true regression coefficients and
the $ \varepsilon_{i} $'s are independent and identically
distributed random variables whose common distribution may take a
parametric form, or may be unspecified, with zero mean and bounded
variance.

\subsection{Weighted least square estimation for AFT model}
The AFT model cannot be solved using ordinary least squares (OLS)
because it cannot handle censored data. The way to handle censored
data is to introduce weighted least squares method, where weights
are used to account for censoring in the least square criterion.
In particular, Kaplan--Meier weights are used to account for the
censoring. This yields a weighted least squares loss function. The
simple form of the loss function makes this estimation approach
especially suitable for high dimensional data. There are many
studies where weighted least squares is used for AFT models
[\cite{huang2006regularized}, \cite{Kha:13:Variable},
\cite{khan2013variable}, \cite{khan2014variable}]. The SWLS method
gives estimate $ \hat{\theta} = (\hat{\beta_{0}},\hat{\beta})$ of
$ \theta = (\beta_{0},\beta) $ which is defined by
\begin{equation}\label{eq:swls}
\hat{\theta} = \arg\min\limits_{\theta} \left[ \frac{1}{2} \sum_{i
= 1}^{n} w_{i}\left( Y_{(i)} - \beta_{0} - X_{(i)}^{T}\beta
\right)^2\right],
\end{equation}
where $ w_{i} $'s are the Kaplan\textendash Meier weights. Let the
data consist of $ (T_{i}^{*}, \delta_{i},X_{i})$, $ (i = 1, \cdots
, n) $, where $ t_{i}^{*} = \min\,(t_{i}, c_{i}) $, $ \delta_{i} =
I(t_{i} \leq  c_{i}) $ and $ t_{i} $ and $ c_{i} $ represent the
realization of the random variables $ T_{i} $ and $ C_{i} $
respectively. Then K\textendash M weights $ w_{i} $'s are defined
as follows
%
%Let the ordered failure and censoring times be $ t_{j}^{*}\,(j = 1, \cdots , n) $, $ d_{j} $ be the number of individuals who fail at time $ t_{j} $ and $ e_{j} $ be the number of individuals censored at time $ t_{j} $. Then the K\textendash M estimate of $ S(t) = P(T_{i} > t) $ is defined as
%
%\begin{equation}
%\hat S(t) = \prod_{\left\lbrace j:t_{j}\le t\right\rbrace }^{} \left( 1 - \frac{d_{j}}{r_{j}} \right) ,
%\end{equation}
%where $ r_{j} = \sum_{i = 1}^{n} I(t_{j} \ge t) $ is the number of individuals at risk at time $ t $. In Stute (1993, 1996)

\begin{equation}
w_{1} = \frac{\delta_{(1)}}{n}, ~ w_{i} =
\frac{\delta_{(i)}}{n-i+1} \prod_{j=1}^{i-1} \left(
\frac{n-j}{n-j+1} \right)^{\delta_{(j)}},
\end{equation}
where $i = 2, \cdots , n$.
%
%
%
%In matrix notation the objective function of SWLS (\ref{eq:swls}) is given by
%
%\begin{equation}\label{eq:swlso}
%\frac{1}{2} \left( Y -\beta_{0}  - X\beta \right)^{T}w\left( Y - \beta_{0} - X\beta \right),
%\end{equation}
%where $ w $ is the $ n \times n $ diagonal weight matrix.
%
Let the uncensored and censored data be subscripted by $ u $ and $
\bar{u} $ respectively. Thus the number of uncensored and censored
observations are denoted by $ n_{u} $ and $ n_{\bar{u}} $, the
predictor and response observations for censored data by $
X_{\bar{u}} $ and $ Y_{\bar{u}} $, and the unobserved true failure
time for censored observation by $ T_{\bar{u}} $. Then SWLS
objective function in matrix notation can be written as
\begin{equation}\label{eq:swlso2}
L\left(\beta_{0},\beta \right) = \frac{1}{2} \left( Y_{u} -
\beta_{0} - X_{u}\beta \right)^{T}w_{u}\left( Y_{u} - \beta_{0} -
X_{u}\beta \right).
\end{equation}
%where $ \lambda_{0} $ is a positive value that accounts for the penalties of violations of constraints, and $ n $ is included for scaling to match the $ w_{u}$.

For notational convenience, we remove $ \beta_{0} $ by
standardisation of the predictors and response i.e.~by centering $
X_{i} $ and $ Y_{i} $ by their weighted means
\[ \bar{X}_{w} = \frac{\sum_{i=1}^{n}w_{i}X_{(i)}}{\sum_{i=1}^{n}w_{i}}, ~ \bar{Y}_{w} = \frac{\sum_{i=1}^{n}w_{i}Y_{(i)}}{\sum_{i=1}^{n}w_{i}}.\]
Using these weighted centered values the intercept term becomes $
0 $. Then the adjusted predictors and responses are defined by

\[ X_{(i)}^{w} = \left(w_{i} \right)^{\frac{1}{2}}\left(X_{(i)} - \bar{X}_{w} \right), ~  Y_{(i)}^{w} = \left(w_{i} \right)^{\frac{1}{2}}\left(Y_{(i)} - \bar{Y}_{w} \right).\]
For simplicity, we still use $ X_{(i)} $ and $ Y_{(i)} $ to denote
the weighted and centered values and $ (Y_{(i)}, \delta_{(i)},
X_{(i)}) $ to denote the weighted data.

The objective function of SWLS (\ref{eq:swlso2}) therefore becomes
\begin{equation}
L\left(\beta \right) = \frac{1}{2} \left(Y_{u} - X_{u}\beta
\right)^{T}\left(Y_{u} - X_{u}\beta \right).
\end{equation}
So, it is easy to show that the SWLS in Eq.~(\ref{eq:aft}) is
equivalent to the OLS estimator excluding the intercept term on
the weighted data with K\textendash M weights. Unfortunately, OLS
estimation does not perform well with variable selection, and is
simply not defined when $ p > n $. Hence various regularized
methods with various penalties are introduced for data where $ p >
n $. The general framework of regularized WLS objective function with
the centered values is defined by

\begin{equation}\label{wlsr}
L\left(\beta,\lambda \right) = \frac{1}{2}\left(Y_{u} - X_{u}\beta
\right)^{T} \left(Y_{u} - X_{u}\beta \right) + \lambda
~\textnormal{pen} (\beta),
\end{equation}
where $\lambda$ is the (scalar or vector) penalty parameter and
the penalty quantity pen $(\beta) $ is set typically in a way so
that it controls the complexity of the model.

\subsection{Lasso estimation for censored data}
The Lasso \cite{tibshirani1997Lasso} minimizes the residual sum of
squares subject to the sum of the absolute value of the
coefficients being less than a constant. The Lasso constraint
selects variables by shrinking estimated coefficients towards 0.
This leads to coefficients exactly equal to zero and allows a
parsimonious and interpretable model. The Lasso can not be applied
directly to the AFT models because of censoring, but the
regularized WLS (\ref{wlsr}) with the Lasso penalty overcomes this
problem. The Lasso estimator $ \hat{\beta} $ for censored data is
obtained as
\begin{equation}\label{eq:Lasso}
\arg\min_{\beta} \frac{1}{2} \left(Y_{u} - X_{u}\beta \right)^{T}
\left(Y_{u} - X_{u}\beta \right) + \lambda_{1}\sum_{j = 1}^{p}
\left|\beta_{j} \right|,
\end{equation}
where $ \lambda_{1} $ is regularization parameter.
\subsection{Ridge estimation for censored data}
Ridge regression \cite{hoerl1970ridge} is a technique for
analyzing multiple regression data that suffer from
multicollinearity. When multicollinearity occurs, least squares
estimates are unbiased, but their variances are large so they may
be far from the true value.
%By adding a degree of bias to the regression estimates, ridge regression reduces the standard errors. It is hoped that the net effect will be to give estimates that are more reliable. Ridge regression is like least squares but minimizes the residual sum of squares subject to the sum of the square of the coefficients being less than a constant that shrinks the estimated coefficients towards zero.
%
The ridge can not be also applied directly to the AFT models
because of censoring, but the regularized WLS (\ref{wlsr}) with
the ridge penalty overcomes this problem. The ridge estimator
$\hat{\beta}$ for censored data is obtained as

\begin{equation}\label{eq:ridge}
\arg\min_{\beta} \frac{1}{2} \left(Y_{u} - X_{u}\beta \right)^{T}
\left(Y_{u} - X_{u}\beta \right) + \lambda_{2}\beta^{T}\beta,
\end{equation}
where $ \lambda_{2} $ is regularization parameter which controls
the strength of the penalty term.

\subsection{Elastic net estimation for censored data}

The elastic net \cite{zou2005regularization} has proved useful
when analyzing data with very many correlated covariates. Elastic
net is like a stretchable fishing net that retains \textquoteleft
all the big fish\textquoteright.
%The $ \ell_{1} $ part of the penalty for elastic net generates a sparse model. On the other hand, the quadratic part of the penalty removes the limitation on the number of selected variables when $ p>>n $ The quadratic part of the penalty also stabilizes the $ \ell_{1} $ regularization path and shrinks the coefficients of correlated predictors towards each other, allowing them to borrow strength from each other.
The elastic net also can not be applied directly to the AFT models
because of having \textquotedblleft censoring
problem\textquotedblright, but the regularized WLS (\ref{wlsr})
with the elastic net penalty overcomes this problem. The elastic
net estimator $ \hat{\beta} $ for censored data is obtained as
\begin{equation}\label{eq:elasticnet}
\arg\min_{\beta} \frac{1}{2} \left(Y_{u} - X_{u}\beta \right)^{T}
\left(Y_{u} - X_{u}\beta \right) + \lambda_{1}\sum_{j = 1}^{p}
\left|\beta_{j} \right| +  \lambda_{2}\beta^{T}\beta,
\end{equation}
where  $ \lambda_{1}  $ \& $  \lambda_{2} $ are $ \ell_{1}  $ \& $
\ell_{2} $ penalty parameters respectively.

\subsection{Stability selection}
The stability selection is proposed by
\cite{meinshausen2010stability} as a very general technique
designed to enhance and improve the performance of the existing
methods.
%In this section, we review stability selection approach to select relevant covariates.
Let $ \beta $ be a $ p $-dimensional vector, where $ \beta $ is
sparse in the sense that $ s < p $ components are non-zero. In
other words, $ \left| \left| \beta\right| \right|_{0} = s < p $.
Denote the set of non-zero values by $ S = \left\lbrace k:
\beta_{k} \neq 0  \right\rbrace  $ and the set of variables with
vanishing coefficient by $ N = \left\lbrace  k : \beta_{k} = 0
\right\rbrace $. The goal of structure estimation is to infer the
set $ S $ from noisy observations. Thus inferring the set $ S $
from data is the well-studied variable selection problem. For a
generic structure estimation or variable selection technique, we
have a regularization parameter $ \lambda \in \Lambda  \subseteq
R^{+} $ that determines the amount of regularization. This
regularization parameters are the penalty parameter in $ \ell_{1}
$-penalized regression and $ \ell_{2} $-penalized regression. For
every value $ \lambda \in \Lambda $, we obtain a structure
estimate $ \hat{S^{\lambda}} \subseteq \left\lbrace 1, \cdots, p
\right\rbrace $. It is then of interest to determine whether there
exists a $ \lambda \in \Lambda $ such that $ \hat{S^{\lambda}} $
is identical to $S$ with high probability. Steps of stability
selection can be summarized as
\begin{itemize}
\item A subsample of size $[\frac{n}{2}]$ is drawn without replacement
denoted by  $I\subseteq \left \{ 1,\cdots ,n \right \}$. \item A
variable selection algorithm is run on  $I$ and it provides set
$\hat{S}^{\lambda} \left ( I \right )$. \item These steps are done
many times. Therefore, for every set $ K \subseteq \left\lbrace1,
\cdots, p \right\rbrace  $, the probability of being in the
selected set $ \hat{S^{\lambda}}(I) $ is
\begin{equation}
\hat{\prod}^{\lambda}_{K} = P (K \subseteq \hat{S^{\lambda}}(I)).
\end{equation}

$P$  is with respest to finite number of subsampling. \item For a
cutoff $ \pi_{thr} $ with $ 0 < \pi_{thr} < 1 $ and a set of
regularization parameters $ \Lambda $, the set of stable variables
is defined as
\begin{equation}
\hat{S}^{stable} = \left\lbrace k : \max_{\lambda \in \Lambda}
\hat{\prod}_{K}^{\lambda} \ge \pi_{thr} \right\rbrace.
\end{equation}
\item The variables with a high selection probability are called
stable variables and kept. \item The variables with a low
selection probability are disregarded. \item The choice of
threshold $\pi_{thr}$ is arbitrary and generally lies in $(0.6,\,
0.9)$.
\end{itemize}

\section{Simulation Studies}
Simulations are used to assess whether subsampling in combination
with a selection algorithm, namely, the lasso, ridge or elastic
net, improves the stability of the algorithm under low dimensional
($p<n$) and high dimensional $(p>n)$ settings.
\subsection{Simulation: $ ~ p = 20, ~n = 50 $ \& $ 100 $}
%\uppercase\expandafter{\romannumeral 1}: $ ~ p = 20, ~n = 50 $
\subsubsection{Setup}
%\textnormal{\&} $ 100 $}
We consider 20 variables divided into two blocks: the first six
variables have $ \beta $ coefficients equal to 5 (i.e. $ \beta_j=5
$ for $ j \in {1, \cdots, 6})$ and the remaining variables have $
\beta $ coefficients equal to zero (i.e. $ \beta_j=0 $ for $ j \in
{7, \cdots , 20}$). The vector of variables $X$ is generated from
the multivariate uniform distribution. i.e. $X \sim U(0, 1)$ such
that the correlation coefficient between the components of $X$ are
either $r=0$, $r=0.2$ or $r=0.5$. For generating correlated
uniform random variables Cholesky decomposition is used.
We first create the correlation matrix $\mathbf{R}$ where the above pairwise
correlation between the $i$-th and $j$-th components of $X$ is set. Then
multiplying the random covariate matrix $\mathbf{X}$ with the Cholesky decomposition of the correlation
matrix $\mathbf{R}$ gives covariates with the desired correlation structure. To see
whether performance is affected by increasing sample size, we
simulate data of sizes $n=50$ and $n=100$. The survival time is
generated from the true AFT model
\begin{equation}\label{aft3.1}
Y_{i} = \beta_{0} + X^{T }_{i}\beta +
\sigma\varepsilon_{i},\,\,\,\,\,\,\, i = 1, \cdots , n,
\end{equation}
where $\epsilon_{i} \sim N(0, 1) $ and $\sigma=1$. Censoring time is generated
from the uniform distribution so as to produce a $ 30 \% $
censoring rate. The AFT model regularized by the lasso, ridge or
elastic net methods is fit to the censored data both with and
without stability selection. The entire process is repeated for
100 simulation runs.
\subsubsection{Evaluation}
The main objective of this study is to evaluate the usefulness of
the stability selection approach in improving variable selection
for the AFT model regularized by the lasso, ridge and elastic net
methods. For a given regularization method, the evaluation is
performed by computing the probability of selection of each
variable under stability selection and without stability selection
over 100 simulated data sets. If stability selection does indeed
improve variable selection, then the probability should be higher
for variables with nonzero $\beta$ coefficients than for variables
whose $\beta$ coefficients are actually zero. In addition, we
record the rate of false positives (F+) and rate of false
negatives (F-). The smaller the values of F+ and F-, the lower the
error. These error rates are shown in Table~\ref{p = 20} for the
three regularization methods considering with and without
stability selection for different choices of $n$ and $r$.
\subsubsection{Results}
\begin{figure}[h]
\centering
\includegraphics[scale=.5]{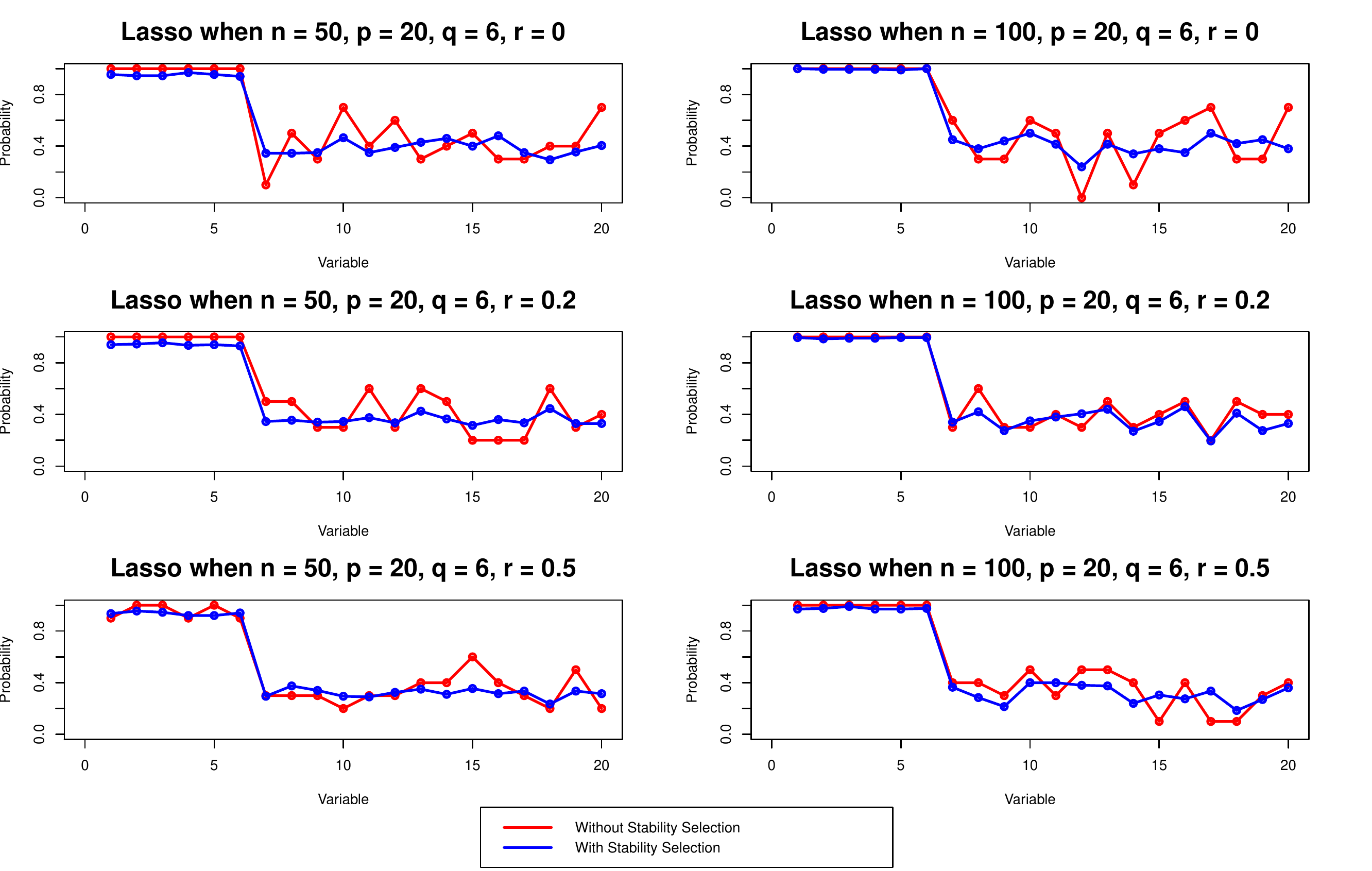}
\caption[Selection Probabilities for Lasso when $ p = 20, ~q = 6 $]{Selection Probabilities for Lasso when $ p = 20, ~q = 6 $}
\label{fig:Lasso-p20-q6}
\end{figure}

\begin{figure}[h]
\centering
\includegraphics[scale=.5]{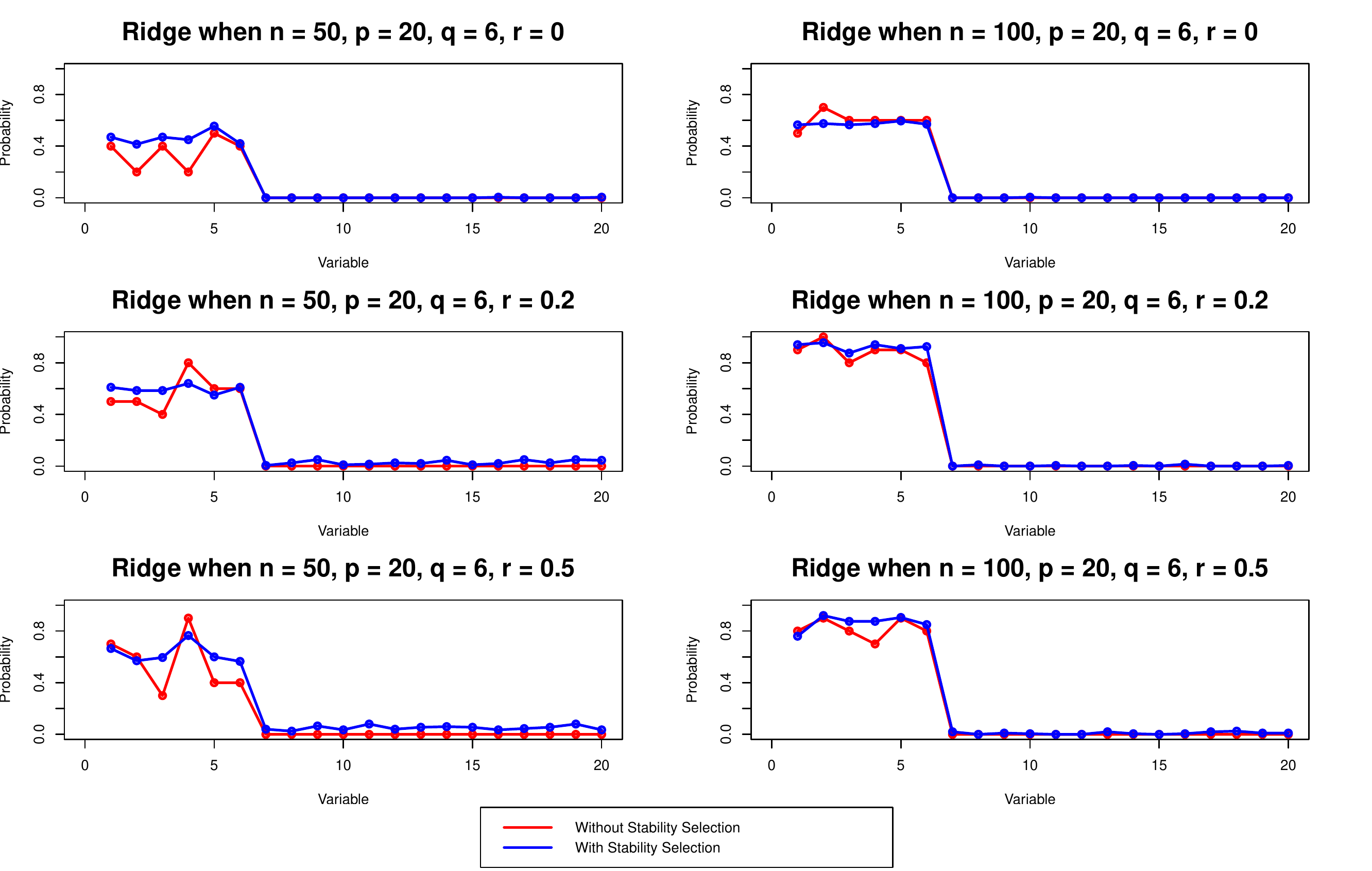}
\caption[Selection Probabilities for Ridge when $ p = 20, ~q = 6 $]{Selection Probabilities for Ridge when $ p = 20, ~q = 6 $}
\label{fig:ridge-p20-q6}
\end{figure}

\begin{figure}[h]
\centering
\includegraphics[scale=.5]{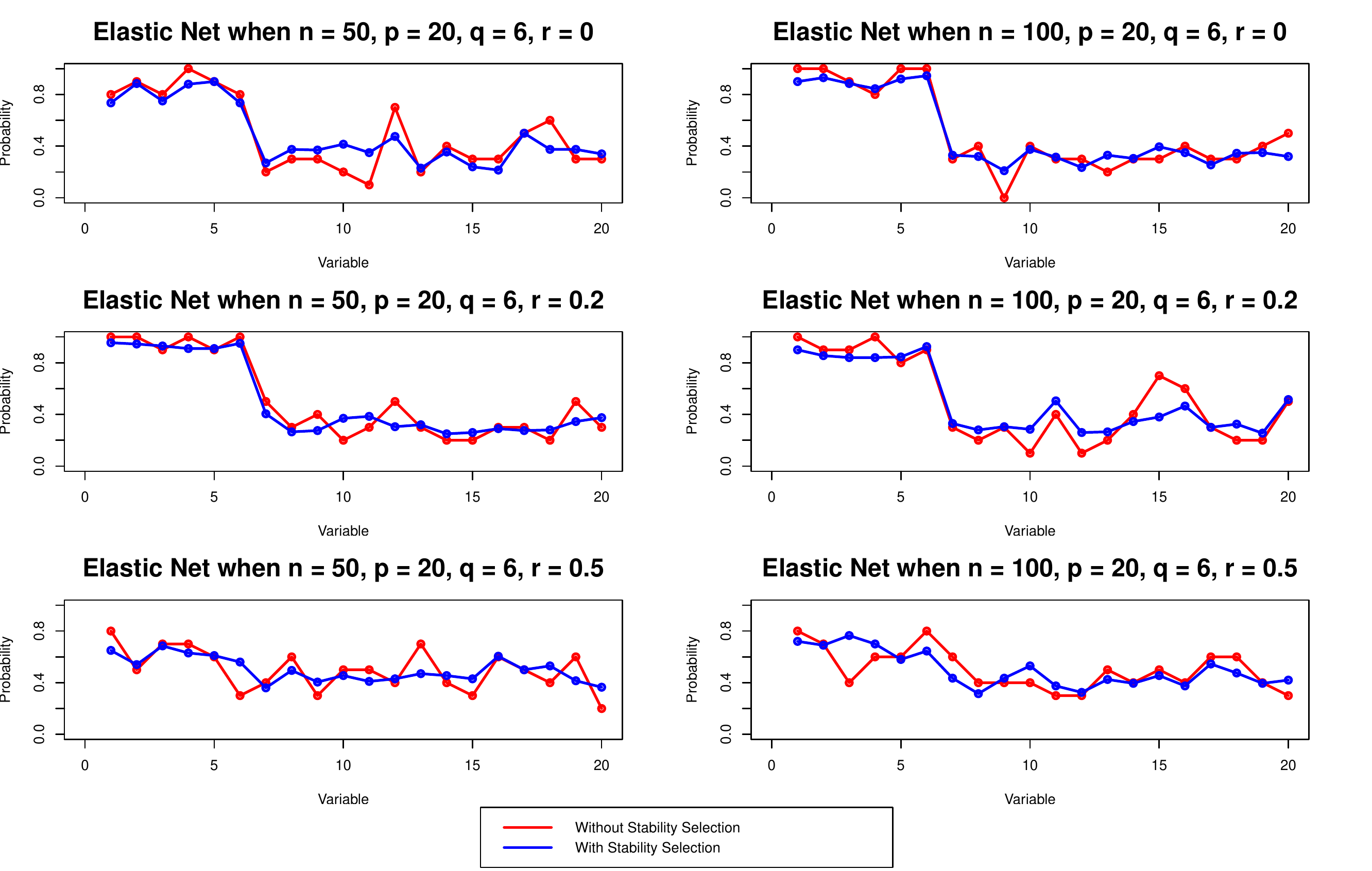}
\caption[Selection Probabilities for Elastic net when $ p = 20, ~q = 6 $]{Selection Probabilities for Elastic net when $ p = 20, ~q = 6 $}
\label{fig:elasticnet-p20-q6}
\end{figure}

From Table~\ref{p = 20} we see that
compared to without stability selection, F+ is smaller when
stability selection is applied to the Lasso and elastic net
methods and this holds true at all sample sizes and correlation
structures. In case of ridge regression, F+ is similar or slightly
higher when stability selection is applied as compared to without
stability selection. Table~\ref{p = 20} also shows that smaller
values for F- are obtained when stability selection is applied to
ridge regression. In case of the Lasso and elastic net, F- is
either slightly higher when stability selection is applied or
similar to the values obtained without stability selection when
$r=0$ or 0.2. An exception occurs when $r=0.5$ in case of the
elastic net where the F- values are lower when stability selection
is applied. The selection probability of each variable under the
regularization methods with stability selection and without
stability selection are shown in Figure~\ref{fig:Lasso-p20-q6} for
Lasso, in Figure~\ref{fig:ridge-p20-q6} for ridge regression and
in Figure~\ref{fig:elasticnet-p20-q6} for elastic net.

According to these figures the selection probabilities of first
six variables are significantly higher than the remaining fourteen
variables with the exception of the elastic net for $r=0.5$. This
is consistent with our simulated data structure in which the first
six variables are designed to have a significant effect on
survival time while the remaining variables have no effect. The
selection probabilities of the variables remain quite stable when
stability selection is applied whereas they tend to fluctuate when
there is no stability selection. This is true irrespective of the
magnitude of the correlation between variables and sample size.
For a conventional choice of the threshold parameter $\pi_{thr}$,
use of the stability selection approach ensures that the
significant variables are selected with high probability. On the
other hand, there is a tendency for false positive errors and
false negative errors to occur when stability selection is not
used.

\subsection{Simulation \uppercase\expandafter{\romannumeral 2}: $ ~ p = 60, ~n = 50 $}
\subsubsection{Setup}
We consider 60 variables among which the first 20 variables are
assigned $ \beta $ coefficients equal to 5 (i.e. $ \beta_j=5 $ for
$ j \in {1, \cdots, 20}$)$(q=20)$ while the remaining are assigned
$ \beta $ coefficients equal to zero (i.e. $ \beta_j=0 $ for~$ j
\in {21, \cdots , 60}) $.  We generate data sets of size $n=50$
where  $ X \sim U(0, 1) $ with correlation between the components
in $X$ being $r=0$, $r=0.2$ or $r=0.5$. This yields the small $n$
large $p$ scenario. All other setting are the same as in
Simulation-$I$.
\subsubsection{Evaluation}
Efficacy of the stability selection approach is evaluated in terms
of the probability of selection for each variable and magnitudes
of the false positive (F+) and false negative (F-) rates computed
over 100 simulated data sets.
\subsubsection{Results} 
\begin{figure}[h]
\centering
\includegraphics[scale=.72]{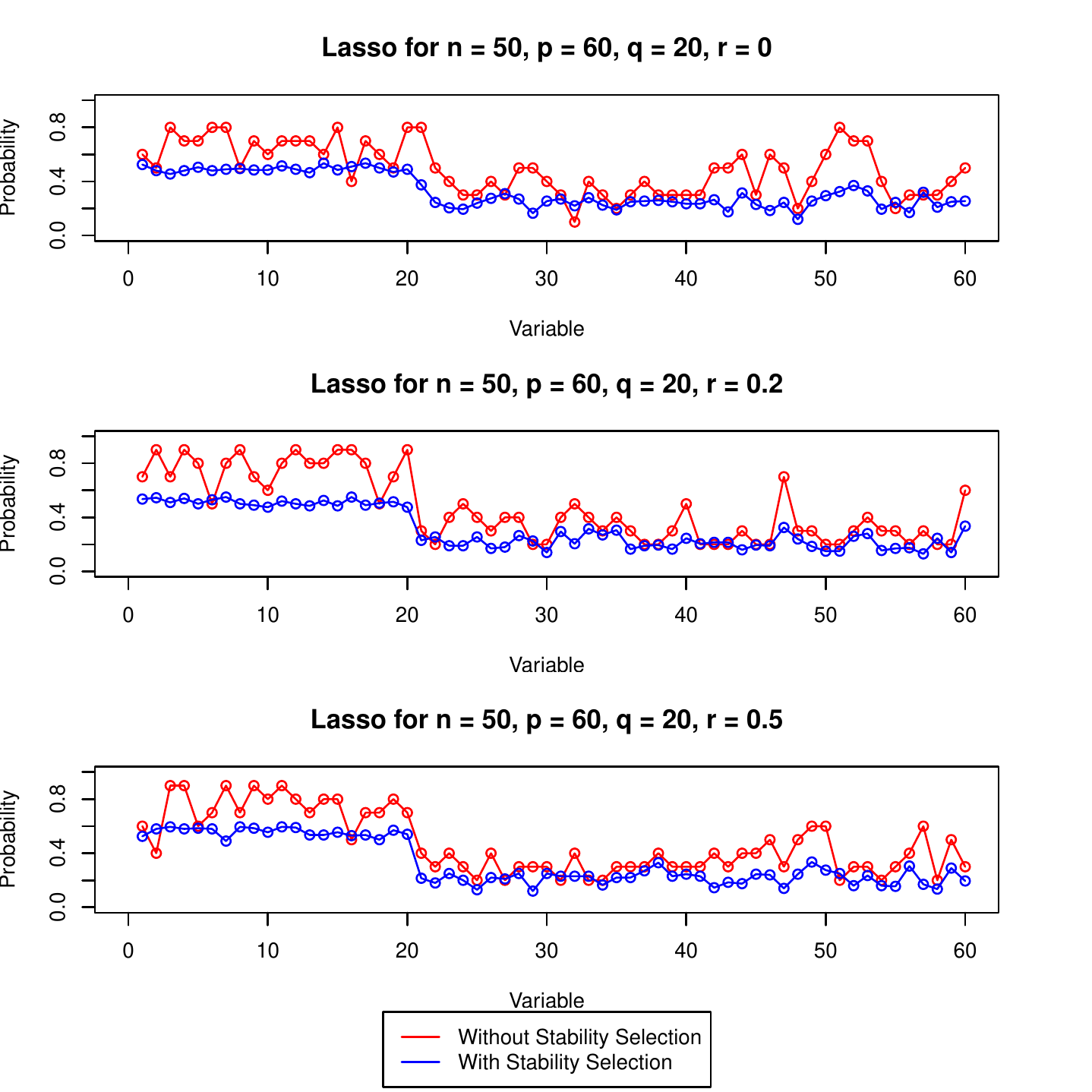}
\caption[Selection Probabilities for Lasso when $ p = 60, ~q = 20 $]{Selection Probabilities for Lasso when $ p = 60, ~q = 20 $}
\label{fig:Lasso-p60-q20}
\end{figure}

\begin{figure}[h]
\centering
\includegraphics[scale=.72]{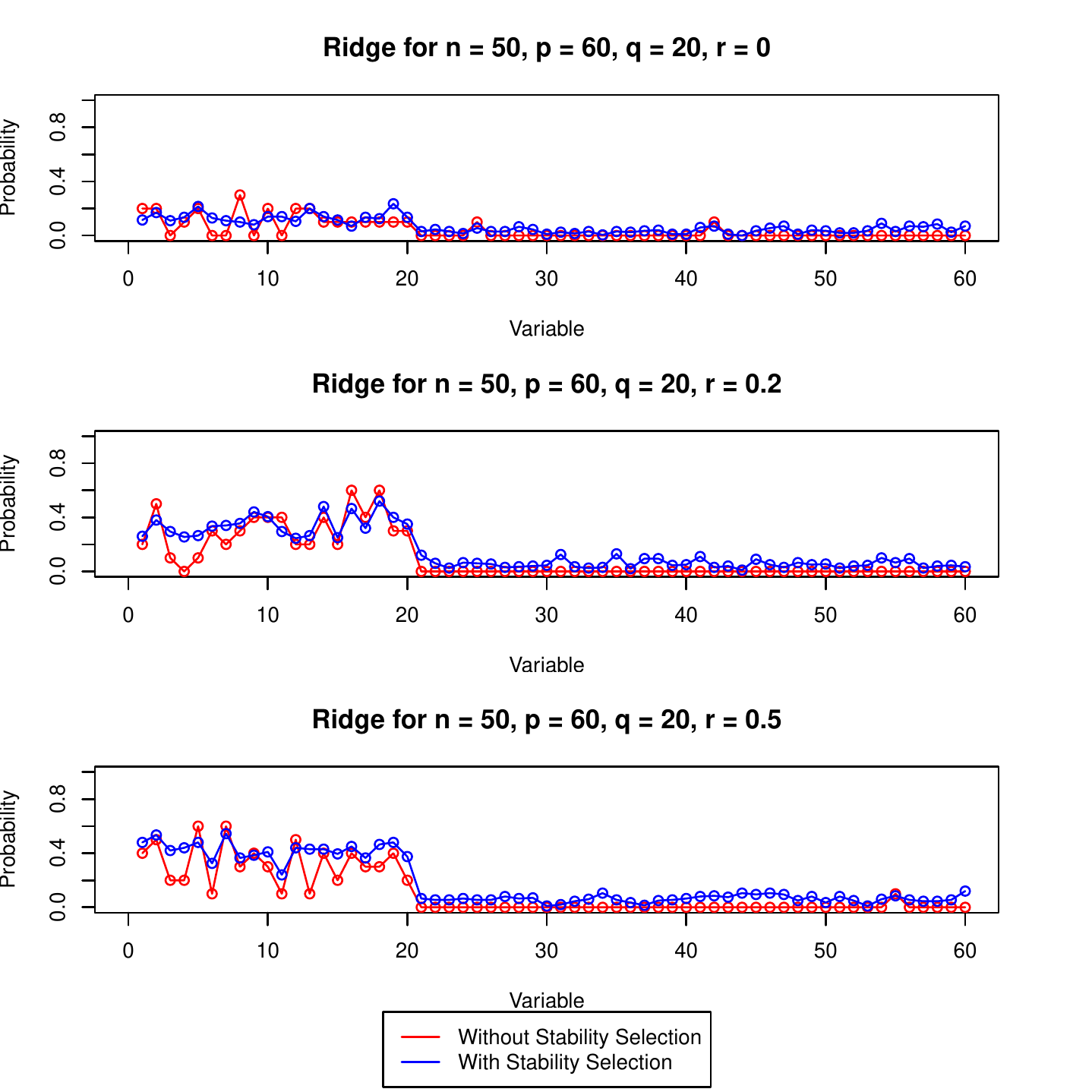}
\caption[Selection Probabilities for Ridge when $ p = 60, ~q = 20 $]{Selection Probabilities for Ridge when $ p = 60, ~q = 20 $}
\label{fig:ridge-p60-q20}
\end{figure}

\begin{figure}[h]
\centering
\includegraphics[scale=.72]{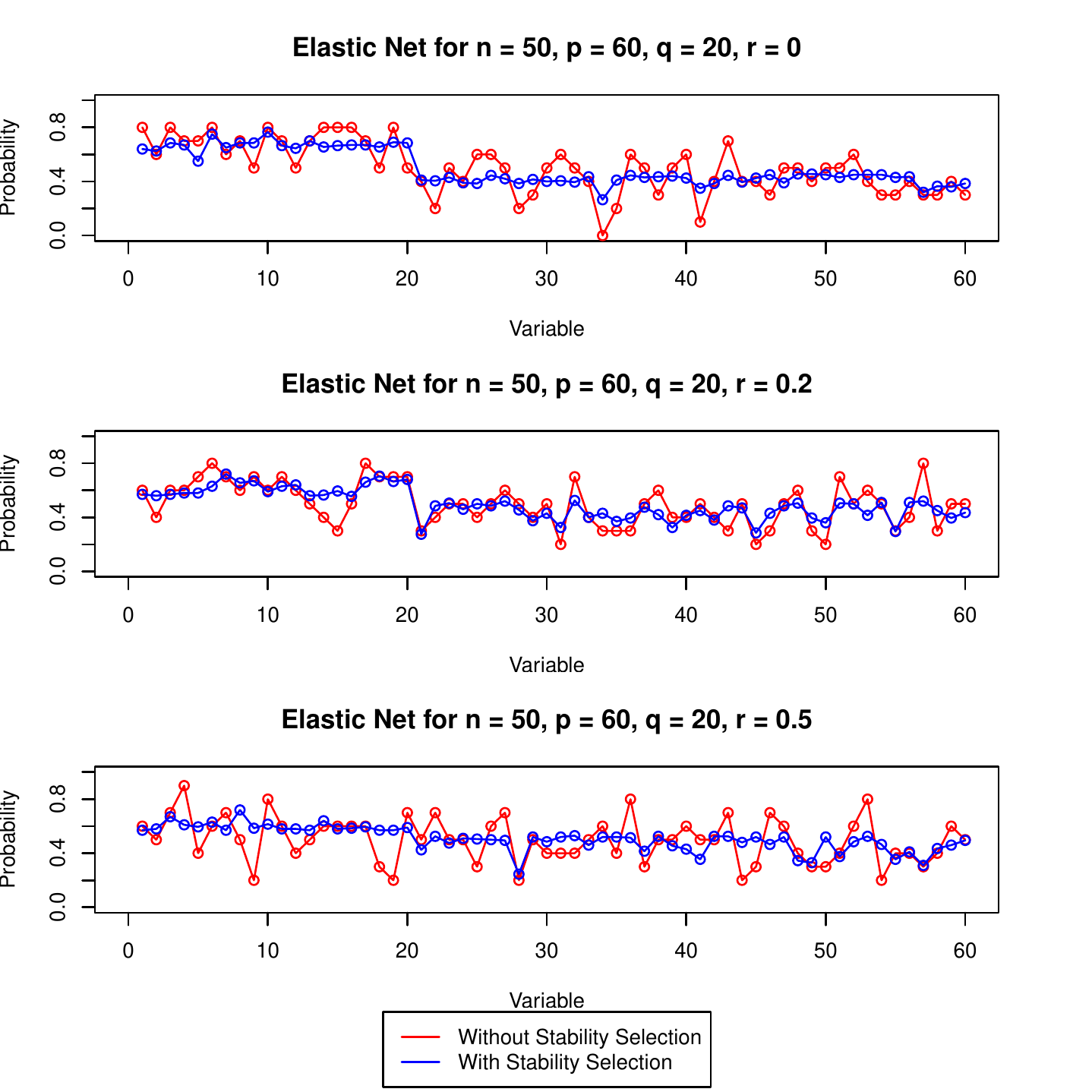}
\caption[Selection Probabilities for Elastic Net when $ p = 60, ~q = 20 $]{Selection Probabilities for Elastic Net when $ p = 60, ~q = 20 $}
\label{fig:elasticnet-p60-q20}
\end{figure}

The error rates are given in Table~\ref{p =
60}. From this table it is seen that for the lasso, F+ is
significantly lower when stability selection is applied whereas
for the other two methods, F+ is nearly the same with or without
stability selection. In general, F- declines for the ridge and
elastic net methods when stability selection is applied, however,
it increases significantly for the lasso method. The advantage of
stability selection in improving variable selection for high
dimensional survival data becomes apparent if one observes
Figures~\ref{fig:Lasso-p60-q20}, ~\ref{fig:ridge-p60-q20} and
~\ref{fig:elasticnet-p60-q20} for the lasso, ridge and elastic
net, respectively. There appears to be large fluctuations in the
selection probability of variables irrespective of the level of
collinearity between variables when stability selection is not
performed. As a result, there is greater error in variable
selection as some of the significant variables have low
probabilities of selection whereas some of the insignificant
variables have relatively large probabilities of selection
resulting in large false positive and false negative rates. In
contrast, the selection probabilities are stable under stability
selection. Furthermore, with the exception of the elastic net at
$r=0.5$, the selection probabilities of the significant variables
are larger than that of the insignificant variables so that errors
are less likely under stability selection. Thus, the results of
simulation-$II$ suggest that for a conventional choice of
threshold parameter \cite{meinshausen2010stability}, $\pi_{thr}$, a regularization method combined
with the stability selection approach can be useful in identifying
significant variables in the high dimensional scenario with
minimum error.

\section{Breast Cancer Data Analysis}
%\subsection{Data description}
We evaluate the stability selection approach for the regularized
AFT model using a breast cancer data set that contains the
metastasis-free survival times from the study of Veer et
al.~\cite{van2002gene}. Here, a series of 295 patients with
primary breast carcinomas were classified as having a
gene-expression signature associated with either a poor or a good
prognosis. We restrict the study to the 144 patients who had lymph
node positive disease and evaluate the predictive value of the
gene-expression profile of patients for the 70 genes previously
determined by Veer et al.~\cite{vant2002gene} based on a
supervised learning method. Five clinical risk factors and 70 gene
expression measurements that were found to be prognostic for
metastasis-free survival are recorded. The censoring rate is 66\%.
The variables in the data set are:
\begin{itemize}
\item \texttt{time:} metastasis-free follow-up time, \item
\texttt{event:} censoring indicator (1 = metastasis or death; 0 =
censored), \item \texttt{diam:} diameter of the tumor (two
levels), \item \texttt{N:} number of affected lymph nodes (two
levels), \item \texttt{ER:} estrogen receptor status (two levels),
\item \texttt{grade:} grade of the tumor (three ordered levels),
\item \texttt{age:} age of the patient at diagnosis, \item
\texttt{TSPYL5 $ \cdots $ C20orf46:} gene expression measurements
of 70 prognostic genes.
\end{itemize}

%\subsection {Previous findings on breast cancer data set}
Walschaerts et al.~\cite{walschaerts2012stable} analyzed this
breast cancer data set in their study which dealt with stable
variable selection methodology. They focussed on new stable
variable selection methods based on bootstrap with two survival
models--the Cox proportional hazard model and survival trees. Cox
model was implemented with two variable selection
techniques--bootstrap Lasso selection (BLS) and Bootstrap
randomized Lasso selection (BRLS). The survival tree was
implemented with the two variable selection techniques--bootstrap
node-level stabilization (BNLS) and Random Survival Forests (RSF).

Six covariates were selected by BLS whereas only four were
selected by the BRLS procedures in a particular setting for those
methods. Six selected covariates by BLS are: \texttt{PRC1},
\texttt{QSCN6L1}, \texttt{QSCN6L1}, \texttt{IGFBP5.1},
\texttt{ZNF533}, \texttt{COL4A2}, \texttt{Contig63649\_RC}. Four
selected variables by BRLS are: \texttt{IGFBP5.1}, \texttt{PRC1},
\texttt{ZNF533}, \texttt{QSCN6L1}. Tree based procedure RSF and
BNLS showed that the most important variable is \texttt{ZNF533}.
BNLS selected three other covariates i.e.~\texttt{COL4A2, PRC1}
and \texttt{N} and the six most important variables selected by
RSF are: \texttt{ZNF533, PRC1, Age, COL4A2, IGFBP5.1, N}. They
found that RSF performed the best for prediction because it
produced the lowest prediction error rate and selected the most
relevant variables to explain the survival durations.

\subsection{Lasso, ridge and elastic net with stability selection}
Applying stability selection approach on Lasso, we find selection
probabilities of all variables. Among them, ten variables have
selection probabilities greater than a threshold value of 0.6:
\texttt{Age(0.955)}, \texttt{ZNF533(0.845)}, \texttt{SCUBE2
(0.8)}, \texttt{N(0.75)}, \texttt{Grade(0.70)},
\texttt{PRC1(0.70)}, \texttt{GPR180 (0.675)}, \texttt{Contig 63649
RC (0.645)}, \texttt{IGFBP5.1(0.615)}, \texttt{COL4A2 (0.610)}.
Thus, the results obtained from stability selection for Lasso is
close enough to the findings in the study by
\cite{walschaerts2012stable}. Lasso with stability selection
selects the same variables as RSF in \cite{walschaerts2012stable}.

Applying stability selection approach on ridge regression, we find
seven variables which have selection probabilities greater than
0.6: \texttt{PRC1(0.805)}, \texttt{COL4A2 (0.715)},
\texttt{IGFBP5.1 (0.625)}, \texttt{GPR180 (0.675)}, \texttt{Contig
63649 RC (0.670)}, \texttt{ZNF} \texttt{533} \texttt{(0.65)},
\texttt{PALM2.AKAP2 (0.615)}. Thus, ridge under stability
selection is also able to select seven variables of which five
were selected by the RSF \cite{walschaerts2012stable}.

Applying stability selection approach on elastic net, we find ten
variables which have selection probabilities greater than 0.6:
\texttt{Age(0.90)}, \texttt{N(0.85)}, \texttt{ZNF533}
\texttt{(0.84)}, \texttt{Grade(0.75)}, \texttt{SCUBE2(0.70)},
\texttt{PRC1(0.65)}, \texttt{IGFBP5.1(0.65)}, \texttt{COL4A2}
\texttt{(0.630)}, \texttt{Contig63649\_RC(0.60)}, \texttt{CENPA(0.60)}. The performance of elastic net under
stability selection is also close to the RSF method
\cite{walschaerts2012stable}. Thus, the three regularization
methods when combined with stability selection yield the most
relevant variables that affect survival. Furthermore, the results
are consistent with the findings of \cite{walschaerts2012stable}.

\subsection{Lasso, ridge regression and elastic net without stability selection}

Applying the AFT regularized by the Lasso on the breast cancer
data without stability selection yields fifteen variables:
\texttt{N}, \texttt{Grade},\texttt{ Age}, \texttt{QSCN6L1},
\texttt{SCUBE2}, \texttt{GMPS}, \texttt{GPR180}, \texttt{ZNF533},
\texttt{RTN4RL1}, \texttt{Contig 63649\_RC}, \texttt{SLC2A3},
\texttt{HRASLS}, \texttt{PALM2.AKAP2}, \texttt{PRC1},
\texttt{ESM1}. Among these variables (\texttt{COL4A2},
\texttt{IGFBP5.1}) were not found significant in the study by
\cite{walschaerts2012stable}. When ridge regression is applied,
fifty five variables are found to have high estimated coefficients
(e.g.~greater than absolute value of one). Although a large number
of variables are selected by ridge regression, two important
variables (Age, \texttt{IGFBP5.1}) that were found significant in
\cite{walschaerts2012stable} were not selected. When the elastic
net was applied, nineteen variables were found to be significant:
\texttt{Diam}, \texttt{N}, \texttt{ER}, \texttt{Grade},
\texttt{Age}, \texttt{QSCN6L1}, \texttt{P5.860F19.3},
\texttt{C16orf61}, \texttt{SCUBE2}, \texttt{ECT2}, \texttt{GSTM3},
\texttt{ZNF533}, \texttt{RTN4RL1}, \texttt{TGFB3},
\texttt{IGFBP5}, \texttt{RTN4RL1}, \texttt{IGFBP5.1},
\texttt{CENPA}, \texttt{ NM\_004702}. However, two variables
(\texttt{COL4A2, PRC1}) that were found to be important in the
study by \cite{walschaerts2012stable} using the RSF method were
not selected. Thus, regularization of the AFT without stability
selection failed to yield a parsimonious model and omitted
variables that were found to be important by previous studies.

\section{Discussion}
Stability in the selection of variables is an important issue when
analyzing high dimensional data. Stability selection methods have
been evaluated in the context of linear regression and more
recently for Cox regression \cite{walschaerts2012stable}. This study
has evaluated whether the stability selection method proposed by
\cite{meinshausen2010stability} can be used to improve variable
selection in the AFT model regularized by the lasso, ridge and
elastic net methods. These evaluations were made through
simulations conducted across different scenarios. False
discoveries are a reason for concern in biomedical research since
they reduce the reliability of results. In the low dimensional
setting ($p<n$), false positive rates were found to be lower when
stability selection was applied for most of the cases. Even in the
high dimensional scenario ($p>n$), false positive rates decreased
in general when stability selection was applied. The selection
probabilities were stable for both high and low dimensional
survival data when stability selection was applied whereas there
were large fluctuations in these probabilities when stability
selection was not applied. As a result, some of the significant
variables had low selection probabilities while some of the
nonsignificant variables had large selection probabilities which
increased the likelihood of observing larger numbers of false
positives and false negatives. Overall, the advantage of stability
selection was apparent across different sample sizes and
correlations between variables.

Analysis of the breast cancer data using regularization methods
combined with stability selection yielded a parsimonious model
containing a small number of variables that were found to be
important by other studies as well
e.g.~\cite{walschaerts2012stable} and \cite{vant2002gene}. In
contrast, a very large number of variables were selected without
stability selection, some of which, were less important and
therefore contributed to false positives. On the other hand, some
variables found relevant by other studies were not chosen due to
their low selection probabilities. Thus, F+ and F- are both likely
to be large without stability selection irrespective of the
regularization method used. In short, we can conclude that the
performance of regularization methods improves when combined with
stability selection, particularly, for high--dimensional censored
data even when there is collinearity between the covariates
leading to a more parsimonious model.

\section*{Conflict of interest statement:}	
The authors have declared no conflict of interest.

%
%%\lhead{}
%%\nocite{khan2013survival}
%\bibliography{Thesis}
%\bibliographystyle{plainnat}

%\section*{Acknowledgements}
%The first author is grateful to the centre for research in Statistical Methodology (CRiSM), Department of Statistics, University of Warwick, UK for offering research funding for his PhD study.
% AOS,AOAS: If there are supplements please fill:
%\begin{supplement}[id=suppA]
%  \sname{Supplement A}
%  \stitle{Title}
%  \slink[doi]{10.1214/00-AOASXXXXSUPP}
%  \sdatatype{.pdf}"
%  \sdescription{Some text}
%\end{supplement}
\bibliography{Thesis}
%\bibliography{reference}
%\bibliographystyle{plain}
%\bibliographystyle{apalike}
\bibliographystyle{wileyj}

%\bibliographystyle{imsart-nameyear}
%\appendix
%\section{Additional Graphs}\label{app}

\begin{table}[h]
\caption{False positive and False negative rate for three methods with stability selection and without stability selection from 100 simulation runs when $ p=20, ~n = 50 $ \& $ 100 $}
\label{p = 20}
\begin{center}
\scalebox{0.70}{
\begin{tabular}{l c c c c c c c c c c c c c c} \toprule

\multirow{4}{*}{Methods} & \multicolumn{4}{c}{r = 0} && \multicolumn{4}{c}{r = 0.2} && \multicolumn{4}{c}{r = 0.5}     \\ \cmidrule(r){2-5} \cmidrule(r){7-10} \cmidrule(r){12-15}
  & \multicolumn{2}{c}{n = 50}   & \multicolumn{2}{c}{n = 100}  && \multicolumn{2}{c}{n = 50}  & \multicolumn{2}{c}{n = 100}  && \multicolumn{2}{c}{n = 50}  & \multicolumn{2}{c}{n = 100}  \\ \cmidrule(r){2-3} \cmidrule(r){4-5} \cmidrule(r){7-8} \cmidrule(r){9-10} \cmidrule(r){12-13} \cmidrule(r){14-15}
  & F+ & F- & F+ & F- && F+ & F- & F+ & F- && F+ & F- & F+ & F- \\ \midrule
\addlinespace
 Lasso without Stability & 0.42 & 0.00 & 0.43 & 0 .00&& 0.39 & 0.01 & 0.38 & 0.00 && 0.34 & 0.05 & 0.34 & 0.00 \\
 Lasso with Stability & 0.38 & 0.04 & 0.40 & 0.00 && 0.35 & 0.05 & 0.34 & 0.00 && 0.30 & 0.06 & 0.31 & 0.02 \\
\addlinespace
 Ridge without Stability & 0.00 &0.65  & 0.00 &  0.45 && 0.00 & 0.43  &0.00  & 0.12 &&  0.00 & 0.45 & 0.00 & 0.19 \\
 Ridge with Stability &0.00  &0.53  &  0.00 & 0.42 &&  0.03 & 0.40 & 0.00 & 0.05 &&  0.05 & 0.38  &  0.00 & 0.14 \\
\addlinespace
 Elastic Net without Stability & 0.34 & 0.13 & 0.32 & 0.05 && 0.32 & 0.03 & 0.32 & 0.10 && 0.46 & 0.40 & 0.44 & 0.35 \\
 Elastic Net with Stability & 0.32 & 0.18 & 0.31 & 0.08 && 0.31 & 0.06 & 0.31 & 0.09 && 0.45 & 0.38 & 0.42 & 0.31 \\
\bottomrule
\end{tabular}
}
\end{center}
\end{table}

%\begin{figure}[ht]
%\centering
%\includegraphics[scale=0.4]{lasso-p20-q6.pdf}
%%\includegraphics[scale=0.4]{simex2.ln.cc.3050CperB.eps}
%%\caption{Predicted vs observed log survival time under log-normal AFT model for the methods AEnet and WEnet for datasets with $P_{\%}= 30$ (first row panel) and $P_{\%}= 50$ (second row panel) and for the methods AEnetCC and WEnetCC for datasets with $P_{\%}= 30$ (third row panel) and $P_{\%}= 50$ (fourth row panel)}.
%\label{fig:simex2.ln.3050}
%\end{figure}

\begin{table}[h]
\caption{False positive and False negative rate for three methods with stability selection and without stability selection from 100 simulation runs for $p = 60, ~n = 50$}
\label{p = 60}
\begin{center}
\scalebox{0.9}{
\begin{tabular}{lccccccccc}
\toprule
\multirow{2}{*}{Methods} && \multicolumn{2}{c}{r = 0}  && \multicolumn{2}{c}{r = 0.2}  && \multicolumn{2}{c}{r = 0.5}   \\ \cmidrule(r){3-4} \cmidrule(r){6-7} \cmidrule(r){9-10}
 && F+ & F- && F+ & F- && F+ & F- \\ \midrule
\addlinespace
Lasso without Stability Selection && 0.41 & 0.34 && 0.32 & 0.23 && 0.34 & 0.26 \\
Lasso with Stability Selection && 0.24 & 0.50 && 0.21 & 0.48 && 0.21 & 0.44 \\
\addlinespace
Ridge without Stability Selection && 0.00 & 0.88 && 0.00 & 0.69  && 0.00  & 0.67 \\
Ridge with Stability Selection && 0.03 & 0.85 && 0.05 & 0.66  && 0.06  & 0.57 \\
\addlinespace
Elastic without Stability Selection && 0.41 & 0.31 && 0.44 & 0.39 && 0.48 & 0.45  \\
Elastic with Stability Selection && 0.41 & 0.32  && 0.43 & 0.38 && 0.47 & 0.40 \\
\bottomrule
\end{tabular}
}
\end{center}
\end{table}

\end{document}